\documentclass[journal]{IEEEtran}

\usepackage{xcolor,soul,framed} 

\colorlet{shadecolor}{yellow}
\usepackage[pdftex]{graphicx}
\graphicspath{{../pdf/}{../jpeg/}}
\DeclareGraphicsExtensions{.pdf,.jpeg,.png}
\usepackage{amsfonts}
\usepackage{multirow}
\usepackage{multicol}
\usepackage{soul}
\usepackage{etoolbox}

\usepackage[utf8x]{inputenc}

\usepackage[cmex10]{amsmath}
\usepackage{amssymb}
\usepackage{array}
\usepackage{mdwmath}
\usepackage{mdwtab}
\usepackage{eqparbox}
\usepackage{url,balance,comment}

\usepackage{siunitx}

\usepackage{tikz}
\usepackage{tikzscale}
\usepackage{graphicx}
\usepackage{tkz-euclide} 
\usepackage{pgfplots,pgfplotstable}
\usetikzlibrary{arrows.meta}
\ifCLASSOPTIONcompsoc
\usepackage[caption=false,font=normalsize,labelfon
t=sf,textfont=sf,subrefformat=parens,labelformat=parens]{subfig}
\else
\usepackage[caption=false,font=footnotesize,subrefformat=parens,labelformat=parens]{subfig}
\fi

\usetikzlibrary{calc,positioning,shapes}

\tikzset{>=latex}
\tikzstyle{diamond marker}=[mark=diamond*, mark options={solid, fill=white, mark size=2.0pt}]
\tikzstyle{triangle marker}=[mark=triangle*, mark options={solid, fill=white, mark size=2.0pt}]
\tikzstyle{square marker}=[mark=square*, mark options={solid, fill=white, mark size=1.3pt}]
\tikzstyle{circle marker}=[mark=*, mark options={solid, fill=white, mark size=1.5pt}]
\tikzset{%
	partial ellipse/.style args={#1:#2:#3}{%
		insert path={+ (#1:#3) arc (#1:#2:#3)}%
	}%
}%

\usepackage{color}

\usepackage[noadjust]{cite}

\usepackage{pgfplots}
\usepackage{pgfplotstable}
\pgfplotsset{compat = newest}

\newlength{\hsep}

\newcommand{\bE}{\mathbf{E}}

\newcommand{\bR}{\mathbf{R}}
\newcommand{\bT}{\mathbf{T}}
\newcommand{\bJ}{\mathbf{J}}
\newcommand{\hbD}{\widehat{\mathbf{D}}}
\newcommand{\hbN}{\widehat{\mathbf{N}}}
\newcommand{\bI}{\mathbf{I}}

\newcommand{\ba}{\pmb{\alpha}}

\newcommand{\psig}{\pmb{\sigma}}
\newcommand{\pdo}[2]{\dfrac{\partial #1}{\partial #2}}
\newcommand{\pdt}[2]{\dfrac{\partial^2 #1}{\partial #2^2}}

\newcommand{\vect}[1]{\ensuremath{\boldsymbol{#1}}}
\newcommand{\mat}[1]{\ensuremath{\mathbf{#1}}}
\newcommand{\imag}{\jmath}

\definecolor{DarkGreen}{rgb}{0.0, 0.5, 0.0}

\newcommand{\Removed}[1]{$\!\!$}

\newcommand{\NoRev}[1]{%
	{
	#1%
	}%
}%

\newcommand{\RevA}[1]{%
	{
	#1%
	}%
}%

\newcommand{\RevB}[1]{%
	{
	#1%
	}%
}%

\graphicspath{{figures/}}

\begin{document}
\bstctlcite{IEEEexample:BSTcontrol}
\title{Revisiting Efficient Multi-Step Nonlinearity Compensation with Machine Learning: An Experimental Demonstration}


  \author{
  Vin\'icius Oliari,
  Sebastiaan Goossens,
  Christian H\"ager,
  Gabriele Liga,
  Rick M.~B\"utler, \\
  Menno van den Hout,
  Sjoerd van der Heide,
  Henry D.~Pfister,
  Chigo Okonkwo,
  and Alex Alvarado
\thanks{%
    V.~Oliari, R.~M.~B\"utler, G.~Liga, S.~Goossens, and A.~Alvarado are with the Information and Communication Theory Lab, Signal Processing Systems Group, Department of Electrical Engineering, Eindhoven University of Technology, Eindhoven, The Netherlands (e-mails: \{v.oliari.couto.dias, r.m.butler, g.liga, s.a.r.goossens, a.alvarado\}@tue.nl).
    
    C.~H\"ager is with the Department of Electrical Engineering, Chalmers University of Technology, Gothenburg, Sweden (e-mail: christian.haeger@chalmers.se).
    
    H.~D.~Pfister is with the Department of Electrical and Computer
	Engineering, Duke University, Durham, USA (e-mail: henry.pfister@duke.edu).
    
    M.~van~den~Hout, S.~van~der~Heide and C.~Okonkwo are with the High Capacity Optical Transmission Lab, Electro-Optical Communication Group, Department of Electrical Engineering, Eindhoven University of Technology, Eindhoven, The Netherlands (emails: \{m.v.d.hout, s.p.v.d.heide, c.m.okonkwo\}@tue.nl).
}
}  

\IEEEspecialpapernotice{(Invited Paper)}

\maketitle

\begin{abstract}
    Efficient nonlinearity compensation in fiber-optic communication systems is considered a key element to go beyond the ``capacity crunch''. One guiding principle for previous work on the design of \emph{practical} nonlinearity compensation schemes is that fewer steps lead to better systems. In this paper, we challenge this assumption and show how to carefully design multi-step approaches that provide better performance--complexity trade-offs than their few-step counterparts. We consider the recently proposed learned digital backpropagation (LDBP) approach, where the linear steps in the split-step method are re-interpreted as general linear functions, similar to the weight matrices in a deep neural network. Our main contribution lies in an experimental demonstration of this approach for a $25$ Gbaud single-channel optical transmission system. It is shown how LDBP can be integrated into a coherent receiver DSP chain and successfully trained in the presence of various hardware impairments. Our results show that LDBP with limited complexity can achieve better performance than standard DBP by using very short, but jointly optimized, finite-impulse response filters in each step. This paper also provides an overview of recently proposed extensions of LDBP and we comment on potentially interesting avenues for future work. 
\end{abstract}


\IEEEpeerreviewmaketitle

\section{Introduction} 
\label{sc:intro}


Mitigating fiber nonlinearity is a significant challenge in high-speed fiber-optic communication systems. As transmission power is increased, the nonlinear Kerr effect degrades the system performance, preventing operation at higher transmission rates, as would be expected in a linear system \cite{Agrell2016}. This performance gap motivates the development of nonlinear compensation techniques, whose design is usually based on analytical models for signal propagation in an optical fiber.     


Digital backpropagation (DBP) based on the split-step Fourier method (SSFM) \cite{AGRAWAL201327} theoretically offers ideal compensation of deterministic propagation impairments including nonlinear effects \cite{Li2008, Mateo2008, Ip2008, Millar2010}. The SSFM is arguably the most popular numerical method to solve the nonlinear Schr\"odinger equation (NLSE) and simulate fiber propagation, while DBP essentially reverses the SSFM operators. Other digital techniques for nonlinearity compensation include Volterra series approximations \cite{Peddanarappagari1997, Gao2009a, Liu2012, Guiomar2012} and recursive perturbation approaches \cite{Yan2011, Tao2011, Liang2014, Nakashima2017}. The main challenge for all these nonlinear compensation techniques is to obtain significant performance improvement and a reasonable computational complexity \cite{Cartledge:17}. Indeed, several authors have highlighted the large computational burden associated with a real-time digital signal processing (DSP) implementation and proposed various techniques to reduce the complexity \cite{Ip2008,Du2010, Rafique2011a, Napoli2014, Jarajreh2015, Giacoumidis2015, Secondini2016, Fougstedt2017, Nakashima2017, Haeger2018ofc}. In many of these works, the number of steps (or compensation stages) is used not only to quantify complexity but also as a general measure of the quality for the proposed complexity-reduction method. The resulting message appears to be that fewer steps are better and provide more efficient solutions. 

While previous work has indeed demonstrated that complexity savings are possible by reducing steps \cite{Du2010, Rafique2011a, Secondini2016}, the main purpose of this paper is to
highlight the fact that fewer steps are not more efficient \emph{per
se}. In fact, recent progress in machine learning suggests that deep
computation graphs with many steps (or layers) tend to be more
parameter-efficient than shallow ones using fewer steps \cite{Lin2017}. In this
paper, we illustrate how this insight can be
applied in the context of fiber-nonlinearity compensation in order to
achieve low-complexity and hardware-efficient DBP. The main idea is to fully parameterize the linear steps in the SSFM by regarding them as general linear functions that can be approximated via finite impulse response (FIR) filters. All FIR filters can then be jointly optimized, similar to optimizing the weight matrices in a deep neural network (NN) \cite{Haeger2018ofc, Haeger2018isit}. Complexity is reduced via pruning, i.e., progressively shortening the filters during the optimization procedure \cite{Fougstedt2018ecoc}. This can be seen as a form of model compression, which is commonly used in machine learning to reduce the size of NNs \cite{Lecun1989, Han2016}. We refer to the resulting approach as learned DBP (LDBP) \cite{Haeger2018ofc}. \RevA{The nonlinear steps in LDBP can also be parameterized and jointly optimized together with the linear steps \cite{Haeger2018ofc}. In this paper, we assume for simplicity that the nonlinear steps remain fixed throughout the optimization procedure (similar to a conventional NN activation function).} 


This paper is an extension of \cite{Haeger2019ecoc}, where we provided a tutorial-like introduction to LDBP. LDBP was originally introduced in \cite{Haeger2018ofc} and the novel technical contribution in this paper lies in an experimental validation of this approach for a single-channel optical transmission system. In particular, it is demonstrated that LDBP with limited complexity can outperform standard DBP by using very short, but jointly optimized, FIR filters in each step. \RevB{\Removed{Our results indicate that LDBP can learn to compensate for experimental impairments and other imperfections that are not directly accounted for by standard DBP, thereby highlighting one of the key strengths of data-driven optimization approaches in a practical setting.}} \NoRev{During the review process of this paper, another experimental demonstration of LDBP was published in \cite{sillekens2019experimental}. Besides the different system parameters adopted for the experiments (such as fiber length, symbol rate, and transmitted constellation), our work differs from \cite{sillekens2019experimental} in terms of the methodology followed in the LDBP pre-optimisation stage: whilst we used experimental data to optimize only the LDBP parameters, in \cite{sillekens2019experimental} two MIMO filters are jointly optimized together with LDBP. In another recent work, the authors in \cite{Bitachon2020} propose a new training method for LDBP in the presence of practical impairments such as laser phase noise. Their approach relies on extracting the relevant impairment estimates from a standard DSP chain based on chromatic dispersion (CD) compensation, which is similar to our approach discussed in Sec.~\ref{sc:finet}.}



This paper is organized as follows. In Sec.~\ref{sc:fiberprop}, we review the theoretical background behind optical fiber propagation and DBP.
Sec.~\ref{sc:ldbp} introduces LDBP and shows how machine learning can be applied in the context of fiber nonlinearity compensation.
Sec.~\ref{sc:results} presents the experimental results and the comparison between DBP and LDBP.
Sec.~\ref{sc:outlook} provides a tutorial-style overview of related works and indicates possible avenues for future work. Finally,
Sec.~\ref{sc:conc} concludes the paper.

\section{Background} 
\label{sc:fiberprop}
In this section, we review the mathematical foundation for LDBP. The optical field propagating in a fiber can be represented by a vector function of time $t$ and distance $z$, $\bE(t,z) = [\NoRev{E_\text{x}} \color{black}(t,z), \NoRev{E_\text{y}} \color{black}(t,z)]^{\top}$, which takes values in $\mathbb{C}^2$, where $\NoRev{E_\text{x}} \color{black}$ and $\NoRev{E_\text{y}} \color{black}$ are the components of the optical field over 2 arbitrary orthogonal polarization modes \NoRev{$\text{x}$} and \NoRev{$\text{y}$}. 
The evolution of $\bE$ in a birefringent optical fiber in the presence of polarization-mode dispersion (PMD) is described by the Manakov-PMD equation \cite{Marcuse1997} as
\begin{align}
    \pdo{\bE(t,z)}{z} = & \left(-\dfrac{\ba(z)}{2}   -\dfrac{j \beta_2 }{2}\pdt{}{t} - \Delta \beta '(z)  \overline{\psig}(z) \pdo{}{t} \right) \bE(t,z) \notag \\
    &+j\gamma \dfrac{8}{9}|\bE(t,z)|^2
    \bE(t,z), 
    \label{eq:mnkvpmd}
\end{align}
where $\ba \in \mathbb{R}^{2\times 2}$ models the polarization-dependent attenuation and amplification effects in a fiber link \cite{Ip2010}, $\beta_2$ is the group-velocity dispersion coefficient, $\gamma$ is the nonlinear coefficient, and $ \Delta \beta '$ is the delay per unit length along the 2 local principal states of polarizations whose evolution is described by the matrix\footnote{\RevA{$\text{SU}(2)$ denotes the special unitary group of degree 2.}} \RevA{ $\overline{\pmb{\sigma}}(z) \in \text{SU}(2)$}. When polarization dependent attenuation/amplification effects can be neglected, then $\ba(z)=\alpha \bI$, where $\alpha\in \mathbb{R}$ is the fibre attenuation coefficient and $\bI$ represents the identity matrix.

Although \eqref{eq:mnkvpmd} does not have a known closed-form solution, an approximated solution can be obtained using the Baker–Campbell–Hausdorff formula \cite{Gilmore1974} as
\begin{align}\label{eq:BakerHausdorff}
\begin{split}
    &\bE(t,z+h)\approx \exp{\left(\int_{z}^{z+h}\hbD(\xi)\text{d}\xi\right)}\\
    &\exp{\left(\int_{z}^{z+h}\hbN(\xi)\text{d}\xi\right)}\bE(t,z),
\end{split}
\end{align}
where 
$\hbD$ and $\hbN$ are the so-called linear and nonlinear operators, respectively, given by
\begin{align}\label{eq:linop}
    \hbD(z) &= -\dfrac{\ba(z)}{2}   -\dfrac{j \beta_2 }{2}\pdt{}{t}- \Delta \beta '(z)  \overline{\psig}(z) \pdo{}{t}, \\
    \hbN(z) &= j\gamma \dfrac{8}{9}|\bE(t,z)|^2. \label{eq:nliop}
\end{align}
The error incurred using \eqref{eq:BakerHausdorff} as a solution of \eqref{eq:mnkvpmd} is vanishingly small as $h$ decreases \cite{Sinkin2002}. This approximation is the main idea behind the SSFM, which underpins DBP.

In the SSFM, the linear and nonlinear operators in \eqref{eq:BakerHausdorff} are recursively applied in frequency and time domain, respectively. The exponential of the nonlinear operator in \eqref{eq:nliop} corresponds instead, in the time-domain, to a multiplication by the term
\begin{equation}
\exp\left(j\gamma\frac{8}{9}\int_{z}^{z+h}|\bE(t,\xi)|^2d\xi\right).
\label{eq:nl_op}    
\end{equation}
The exponential of the linear operator $\hbD(\xi)$ in \eqref{eq:linop} can be expressed in the Fourier domain as a (frequency-dependent) matrix multiplication by
\begin{align}
\begin{split}
\exp\left(-\int_{z}^{z+h}\frac{\ba(\xi)}{2}d\xi\right)\exp{\left(j\omega^2\frac{\beta_2}{2} h\right)}\bJ(\omega,z),
\end{split}
\label{eq:linear_op}    
\end{align}
where $\bJ(\omega,z)=\exp\left(-\int_{z}^{z+h}j\omega\Delta \beta '(\xi)\overline{\psig}(\xi)d\xi\right)$ is a unitary, frequency-dependent matrix, commonly referred to as (local) Jones matrix.  
For small enough $h$, the Jones matrix can be factorized as $\bJ(\omega,z)=\bR(z)\bT(\omega,z)$, where $\bR$ is a unitary complex matrix which describes the evolution of the polarization state of the optical field, and where
\begin{align} \label{eq:expdgd}
    \bT(\omega,z) &= \left[ \begin{array}{cc}
        \exp{\left( -j \omega \frac{\tau(z)}{2} \right)} & 0  \\
        0 & \exp{\left( j \omega \frac{\tau(z)}{2} \right)}
    \end{array} \right]
\end{align}
with $\tau(z)=\Delta \beta^{\prime}(z)h$ describes the delay over the two principal states of polarization at section $z$. Together, $\bR(z)$ and $\bT(\omega,z)$ contribute to define the evolution of PMD along the link.

The conventional DBP algorithm aims to reconstruct the transmitted field $\bE(t,0)$ from the received one $\bE(t,z)$ using \eqref{eq:BakerHausdorff} recursively as\footnote{Here, $\prod_{i=1}^N A_i = A_1 A_2 \cdots  A_N$, in the operator product sense.}  
\begin{align}
\begin{split}
\hat{\bE}(t,0)&=\prod_{n=1}^{N_{\text{DBP}}}\exp{\left(-\int_{z_{n}}^{z_{n+1}}\hbD^{\prime}(\xi)\text{d}\xi\right)}\\
    &\exp{\left(-\int_{z_{n}}^{z_{n+1}}\hbN(\xi)\text{d}\xi\right)}\bE(t,z),
\end{split}
\label{eq:DBP}
\end{align}
where $\hbD^{\prime}=-\dfrac{\alpha}{2}-\dfrac{j \beta_2 }{2}\pdt{}{t}$, $z_n$ and $\Delta z_n=z_n-z_{n-1}$ for $n=1,\ldots,N_{\text{DBP}}$ are the propagation section and so-called DBP \emph{step size} at iteration $n$, and $N_{\text{DBP}}$ is the number of DBP steps.
Like in the SSFM implementation, the two operators in \eqref{eq:DBP} are typically applied in frequency- and time-domain for the linear and nonlinear operators, respectively. This is performed numerically via two fast-Fourier transforms (FFTs). 


Comparing \eqref{eq:BakerHausdorff} with \eqref{eq:DBP}, we note that in the conventional DBP implementation, the operator $\hbD^{\prime}$ does not exactly invert $\hbD$ in \eqref{eq:BakerHausdorff}, because \RevA{$\hbD^{\prime}$ does not account for the effects of $\bR(z)$ and $\bT(\omega,z)$. Including these two matrices in $\hbD^{\prime}$ is challenging, since they are stochastically distributed over an ensemble of fibres and unknown to the receiver. Failing to invert $\bR(z)$ and $\bT(\omega,z)$ in a distributed fashion} results in a performance penalty due to the uncompensated interaction between PMD and the Kerr effect \cite{Czegledi2017, Liga2018}. Combining DBP (and LDBP) with distributed PMD compensation is discussed in more detail in Sec.~\ref{sec:pmd}. However, \NoRev{distributed} PMD compensation is not yet integrated into the experimental demonstration. 

\section{Efficient Multi-Step Nonlinearity Compensation using Deep Learning} 
\label{sc:ldbp}

For hardware-efficient and low-complexity DBP, the task is to approximate the solution of the NLSE using as few computational resources as possible. As described in the previous section, the SSFM computes a numerical solution by alternating between linear filtering steps (accounting for CD and attenuation) and nonlinear phase rotation steps (accounting for the optical Kerr effect). It was observed in \cite{Haeger2018ofc} that this is indeed quite similar to the functional form of a deep NN, where linear (or affine) transformations are alternated with pointwise nonlinearities. In this section, we illustrate how this observation can be exploited by applying tools from machine learning, in particular deep learning. 

\subsection{Supervised Learning and Neural Networks}

We start by reviewing the standard supervised learning setting for
feed-forward neural networks (NNs). A feed-forward NN with $M$ layers
defines a mapping $\hat{\vect{y}} = \vect{f}_\theta(\vect{x})$ where the input
vector $\vect{x} \in \mathcal{X}$ is mapped to the output vector
$\hat{\vect{y}} \in \mathcal{Y}$ by alternating between affine
transformations $\vect{z}^{(i)} = \vect{W}^{(i)} \vect{x}^{(i-1)} +
\vect{b}^{(i)}$ and pointwise nonlinearities $\vect{x}^{(i)} =
\phi(\vect{z}^{(i)})$ with $\vect{x}^{(0)} = \vect{x}$ and
$\vect{x}^{(M)} = \hat{\vect{y}}$. The parameter vector $\theta$ comprises
all elements of the weight matrices $\vect{W}^{(1)}, \dots,
\vect{W}^{(M)}$ and vectors $\vect{b}^{(1)},\dots,\vect{b}^{(M)}$.
Given a training set $S\subset \mathcal{X} \times \mathcal{Y}$ that
contains a list of desired input--output pairs, training proceeds by
minimizing the empirical loss $\mathcal{L}_S (\theta) \triangleq
\frac{1}{|S|} \sum_{(\vect{x},\vect{y})\in S} \ell \big(
f_\theta(\vect{x}),\vect{y})$, where $\ell(\hat{\vect{y}},\vect{y})$ is \RevA{a real number that, given a pair $(\vect{x},\vect{y})\in S$, determines the performance of the prediction $\hat{\vect{y}} = f_\theta(\vect{x}) $ when $\vect{y}$ is the correct output target. We call $\ell$ the loss function. In our case, $\vect{x}$ is a vector of received samples after fiber propagation and some impairments compensations, $\vect{y}$ is the vector of transmitted symbols, $\hat{\vect{y}}$ is the estimated symbol vector, and $\ell$ is the mean-squared error (MSE) function $\ell(\hat{\vect{y}},\vect{y}) = \lVert \hat{\vect{y}} - \vect{y} \rVert^2$, where $\lVert \cdot \rVert$ is the Euclidean norm. }
When the training set is
large, one typically optimizes $\theta$ using a variant of stochastic
gradient descent (SGD). In particular, mini-batch SGD uses the
parameter update $\theta_{t+1} = \theta_t - \alpha \nabla_\theta
\mathcal{L}_{B_t} (\theta_t)$, where $\alpha$ is the step size and
$B_t \subseteq S$ is the mini-batch used in the $t$-th step. 

Supervised machine learning is not restricted to NNs and learning algorithms such as SGD can be applied to other function classes as well. In this paper, we do not further consider NNs, but instead focus on approaches where the function $\vect{f}_\theta$ results from parameterizing a model-based algorithm, in particular the SSFM. In fact, prior to the current revolution in machine learning, communication engineers were quite
aware that system parameters (such as filter coefficients) could be
learned using SGD. It was not at all clear, however, that more
complicated parts of the system architecture could be learned as well.
For example, in the linear operating regime, PMD can be compensated by choosing the function
$\vect{f}_\theta$ as the convolution of the received signal with the impulse
response of a linear multiple-input multiple-output (MIMO) filter,
where $\theta$ corresponds to the filter coefficients. For a suitable choice of the loss function $\ell$, applying SGD then maps into the well-known constant modulus algorithm (CMA) \cite{Savory2008}. For the experimental investigation in this paper, the CMA is used as part of our receiver DSP chain as an adaptive equalizer (see Sec.~\ref{sc:dsp}). 

\subsection{Learned Digital Backpropagation}


Real-time DBP based on the SSFM is widely considered to be impractical
due to the complexity of the FFTs commonly
used to implement frequency-domain (FD) CD filtering. To address this
issue, time-domain (TD) filtering with finite impulse response (FIR)
filters has been suggested in, e.g., \cite{Ip2008, Zhu2009,
Goldfarb2009, Fougstedt2017}. In these works, either a single filter
or filter pair is designed and then used repeatedly in each step.
However, using the same filter multiple times is suboptimal in general
and all the filter coefficients used by the DBP algorithm should be
optimized jointly. To that end, it was proposed in \cite{Haeger2018ofc} (see also \cite{Haeger2018isit}) to apply supervised learning based
on SGD by letting the function $\vect{f}_\theta$ be the SSFM, where the
linear steps are now implemented using FIR filters. In this case, $\theta$ corresponds to the filter coefficients used in \emph{all} steps. The resulting method is referred to as LDBP.

\begin{figure*}[t]
    \centering
    \includegraphics[width=0.9\linewidth]{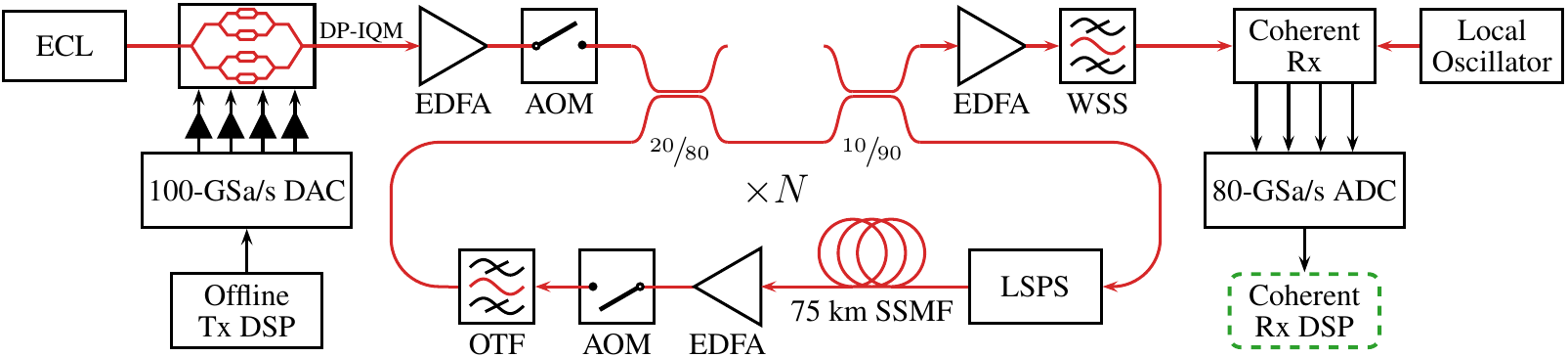}
    \caption{Experimental optical recirculating loop setup (ECL: external cavity laser, DP-IQ Mod: dual polarization IQ Modulator, EDFA: erbium-doped fiber amplifier, AOM: acousto-optic modulator, WSS: wavelength selective switch, DAC: digital-to-analog converter, ADC: analog-to-digital converter, DSP: digital signal processing, OTF: optical tunable filter, LSPS: loop-synchronized polarization scrambler, SSMF: standard single-mode fiber). }
    \label{fig:experimental_setup}
\end{figure*}

\begin{figure*}
    \centering
    \colorlet{color1}{blue!20!white}
\colorlet{color2}{green!30!white}
\colorlet{color3}{red!20!white}
\colorlet{color4}{yellow!20!white}
\definecolor{Cgreen}{RGB}{44 160 44}
\setlength{\hsep}{7ex}

\tikzstyle{Cir} = [draw, circle, 
minimum size=0.5em]

\tikzstyle{Cir} = [draw, circle, 
    minimum size=2em]

\begin{tikzpicture}[font=\normalsize,
>=stealth,auto,
block/.style={rectangle,thick,draw,inner sep=2pt,minimum width=1\hsep, minimum height=1.0\hsep,font=\footnotesize,rounded corners}, 
line1/.style = {draw, thick,->,rounded corners},
amp/.style = {regular polygon,thick, regular polygon sides=3,
              draw, fill=white, text width=1.5em,
              inner sep=0em, outer sep=0mm,
              shape border rotate=-90}]

\coordinate (Fiber); %
\coordinate [left=0.4 of Fiber] (Brack1);
\coordinate [right=2 of Fiber] (AmpC);
\coordinate [right=1 of AmpC] (Brack2);

\coordinate [below left= 1.1 and 0.5 of Brack1] (NspC);
\coordinate [below right = 0.575 and 0.5 of Fiber] (SpanSize);
\coordinate [right=1\hsep of Fiber] (RRCC); 
\coordinate [left=1.5\hsep of RRCC] (QAMC);

\coordinate [right=.7\hsep of Brack2] (BRICKC); 
\coordinate [right=1.4\hsep of BRICKC] (DBPC); 
\coordinate [right=1.5\hsep of DBPC] (MFC); 
\coordinate [right=1.7\hsep of MFC] (QAMDC);

\coordinate [below right = 0.2 and 0.52 of Fiber] (Pol);

\coordinate [below right = 0.74 and 2 of BRICKC] (Samp);
\coordinate [below right = 1.1 and 0.825 of BRICKC] (Samp1);
\coordinate [above right = 0.8 and 0.825 of BRICKC] (Samp2);

\coordinate [below right = 0.74 and 2 of RRCC] (OpD);
\coordinate [below right = 1.1 and 0.69 of RRCC] (OpD1);
\coordinate [above right = 0.8 and 0.69 of RRCC] (OpD2);
\coordinate [below right = -0.025 and 2.575 of OpD1] (OpDT);


\coordinate [right=1\hsep of QAMDC] (RO);
\coordinate [left=1.075\hsep of QAMC] (LO);

\coordinate [above left=1.5\hsep and .85\hsep of QAMC] (R11);
\coordinate [below right=1.25\hsep and .75\hsep of QAMDC] (R12); 
\coordinate [above left=1.125\hsep and .7\hsep of RRCC] (R21);
\coordinate [below right=1.125\hsep and .85\hsep of MFC] (R22); 
\coordinate [above right=1.3\hsep and 0\hsep of Brack2] (PRFBC);
\coordinate [below=.35\hsep of PRFBC] (SNRC);




\node[block,align=center,anchor=west] (QAM) at (QAMC) {Receiver \\ Front End \\ IQ Comp.};

\node[block,align=center,anchor=west] (RRC) at ($(QAM.east)+(0.7,0)$) {Resampling \\ 2 Samp/sym};

\node[block,align=center,anchor=west] (BRICK) at ($(RRC.east)+(0.7,0)$) {Frequency \\ Offset \\ Compensation};

\node[block,align=center,anchor=west, minimum width = 2\hsep] (DBP) at ($(BRICK.east)+(0.7,0)$) {EDC\\DBP\\LDBP};

\node[block,align=center,anchor=west] (QAMD) at ($(DBP.east)+(0.7,0)$) {MIMO Eq. \\ \& Phase Noise \\ Compensation};

\node[block,align=center,anchor=west] (TAS) at ($(QAMD.east)+(0.7,0)$) {Downsampling \&\\ Time Alignment};

\node[block,align=center,anchor=west] (SNRBL) at ($(TAS.east)+(0.7,0)$) {SNR \\ Estimation};


\draw [draw,->,thick] ($(QAM.west)+(-0.5,0)$) -- node [midway,above] {} (QAM);
\draw [draw,->,thick] (QAM) -- node [midway,above] {} (RRC);

\draw [draw,->,thick] (RRC) -- node [midway,above] {} (BRICK);

\draw [draw,thick] (BRICK) -- node [midway,above] {} (DBP);
\draw [draw,->,thick] (DBP) -- node [midway,above] {} (QAMD);
\draw [draw,->,thick] (QAMD) -- node [midway,above] {} (TAS);

\draw [draw,->,thick] (TAS) -- node [midway,above] {} (SNRBL);

\draw [draw,thick] (DBP.west) -- ($(DBP.west)+(0.4,0)$);
\draw [draw,thick,->] ($(DBP.west)+(0.4,0)$) -- ($(DBP.west) + (0.7,0.3)$);
\draw [draw,thick,densely dotted] ($(DBP.west)+(0.4,0)$) -- ($(DBP.west) + (0.75,0.0)$);
\draw [draw,thick,densely dotted] ($(DBP.west)+(0.4,0)$) -- ($(DBP.west) + (0.7,-0.3)$);

\draw [draw,thick] (DBP.east) -- ($(DBP.east)+(-0.4,0)$);
\draw [draw,thick] ($(DBP.east)+(-0.4,0)$) -- ($(DBP.east) + (-0.7,0.3)$);
\draw [draw,thick,densely dotted] ($(DBP.east)+(-0.4,0)$) -- ($(DBP.east) + (-0.75,0.0)$);
\draw [draw,thick,densely dotted] ($(DBP.east)+(-0.4,0)$) -- ($(DBP.east) + (-0.7,-0.3)$);

\draw[Cgreen,dashed,rounded corners,very thick] ($(QAM.west)+(-0.6,0.7)$) rectangle ($(SNRBL.east)+(0.1,-0.7)$);

\node[align=center,above=0.12] at (DBP.north) {{\small Coherent Rx DSP}};







\end{tikzpicture}
    \caption{DSP chain applied to the digitized fiber output detected by the coherent receiver and analog-to-digital converter as shown in Fig.~\ref{fig:experimental_setup}.}
    \label{fig:dspchain}
    \vspace{-0.2cm}
\end{figure*}
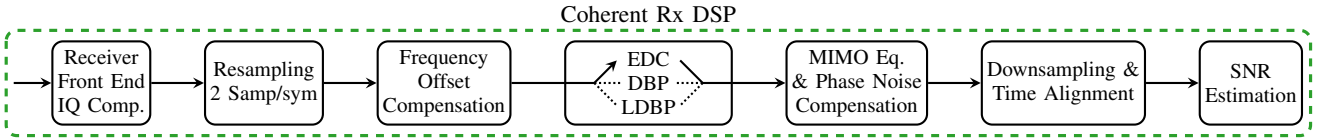

The complexity of LDBP can be reduced by applying \emph{model compression}, which is commonly used in ML to reduce the size of NNs \cite{Lecun1989, Han2016}. In this paper, we use a simple pruning approach where the FIR filters are progressively shortened during SGD \cite{Fougstedt2018ecoc}. Our main finding is that the filters can be pruned to remarkably short lengths without sacrificing performance. As an example, consider single-channel DBP of a $10.7$-Gbaud signal over $25 \times 80$ km of standard single-mode fiber (SSMF) using the SSFM with one step per span (StPS). For this scenario, Ip and Kahn have shown that $70$-tap filters are required to obtain acceptable accuracy
\cite{Ip2008}. This assumes that the filters are designed using FD
sampling and that the same filter is used in each step. The resulting
hardware complexity was estimated to be over $100$ times larger than
for linear equalization. On the other hand, with jointly optimized
filters, it was previously demonstrated that one can achieve similar accuracy by alternating
between filters that are as short as $5$ and $3$ taps
\cite{Haeger2018isit}. This reduces the complexity by almost two
orders of magnitude, making it comparable to linear equalization in this case.


At first glance, it may not be clear why multi-step DBP can benefit from joint optimization of the filters. After all, the standard SSFM applies \emph{the same} CD filter many times in succession, without the need for any elaborate optimization. The explanation is that in the presence of practical imperfections such as finite-length filter truncation, applying the same \emph{imperfect} filter multiple times can be detrimental because it magnifies any weakness. To achieve a good combined response of neighboring filters and a good overall response, the truncation of each filter needs to be delicately balanced. For a more detailed discussion, we refer readers to \cite{Haeger2018isit, Lian2018itw}. 
\section{Experimental Results} 
\label{sc:results}

For the experimental results presented in this section, our focus is on single-channel DBP of a polarization-multiplexed (PM) $25$ Gbaud signal over $1500$ km of SSMF. To obtain the results, we proceed in three steps: 
\begin{enumerate}
    \item Pre-train LDBP using data from split-step simulations and apply filter pruning to obtain short FIR filters in each step. 
    \item Fine-tune the model using pre-processed experimental data traces. 
    \item Test the fully-trained model on raw experimental data traces, by integrating LDBP into the receiver DSP chain. 
\end{enumerate}
\RevA{The pre-processed experimental data traces mentioned in step 2 are obtained using the procedures described in Sec.~\ref{sc:finet}. This pre-processing is necessary in order to provide LDBP with an estimation of effects such as phase noise and state of polarization. The raw experimental data traces in step 3 are the digital domain samples of the fiber output. These raw traces are processed by a different DSP chain, described in Sec.~\ref{sc:dsp}.} In the following, we discuss each step in more detail, starting with the experimental testbed and DSP algorithms used for the experimental validation. Effective SNR is used throughout this section as the main figure of merit. \RevB{We also note that the training procedure described in Sec.~\ref{sc:prefp} and Sec.~\ref{sc:finet} is performed only once, since we only consider static effects, i.e., chromatic dispersion and fiber nonlinearities. Moreover, the trained model is independent of the transmitted power, as described in more detail below. }   

\subsection{Recirculating Loop Setup and DSP Chain}\label{sc:dsp}


A schematic of the experimental recirculating loop setup is depicted in Fig.~\ref{fig:experimental_setup}. \NoRev{A total of 851 traces were captured in the launch power range of $-5$ dBm to $6$ dBm with steps of $0.5$ dBm, accounting for 37 traces per launch power. For each trace, we generate offline a sequence of \(2^{16}\) symbols,} using the Permuted Congruential Generator XSL RR 128/64 random number generator \cite{pcg2014}. \NoRev{A new random seed is used for each sequence.} The sequences are pulse-shaped using a root-raised cosine (RRC) filter with 1\% roll-off, digitally pre-compensated for transmitter bandwidth limitations and uploaded to a 100-GSa/s digital-to-analog converter (DAC). The \(193.4\) THz carrier is generated by an external cavity laser (ECL), modulated by a dual-polarization IQ-modulator (DP-IQM) and amplified using an erbium doped fiber amplifier (EDFA). Using acousto-optic modulators (AOMs), the optical signal is circulated in a recirculating loop consisting of a loop-synchronised polarization scrambler (LSPS), a 75-km span of SSMF, an EDFA and an optical tunable filter (OTF) for gain equalization. 

At the receiver, the optical signal is amplified with an EDFA, filtered using a \RevA{$50$ GHz optical bandwidth} wavelength selective switch (WSS)  and \NoRev{detected using} an intradyne coherent receiver consisting of a local oscillator (LO), 90-degree hybrid and 4 balanced photodiodes \RevA{with $43$ GHz electrical bandwidth}. The resulting electrical signal is digitized by an 80-GSa/s real-time oscilloscope \RevA{with an electrical bandwidth of $36$ GHz.} \NoRev{\Removed{Multiple captures for each transmitted power are taken.}}

The receiver DSP consists of seven blocks, which are applied sequentially as represented in Fig.~\ref{fig:dspchain}. Orthonormalization using blind moment estimation is applied to the signal for receiver optical front-end IQ compensation (gain imbalance and offset angle between the in-phase and quadrature components). Rational resampling to 2 samples per symbol is then applied. The next step is frequency-offset estimation and compensation \RevA{to correct effects such as frequency difference between the local oscillator and the signal laser and frequency offsets introduced by the AOMs}. After frequency-offset compensation, we apply either electronic dispersion compensation (EDC), DBP, or LDBP. The signal is then adaptively equalized and phase noise compensated. Here we use a MIMO equalizer trained with \RevA{MSE} metric. The MIMO equalizer is used to recover the signal state of polarization and partially compensate for other impairments, such as PMD. Within the update loop of the equalizer, blind phase search using the known transmitted symbols removes phase noise. The equalized signal is then downsampled to 1 sample per symbol and aligned with the transmitted sequence. Finally, the effective SNR is estimated.

\subsection{Pre-Training and Filter Pruning}
\label{sc:prefp}

Simulations are used for pre-training where the simulation parameters are closely matched to the experimental setup. In particular, we assume single-channel transmission of a $25$ Gbaud signal (PM $16$-QAM, 1\% RRC) over $20 \times 75.484$ km of fiber ($\alpha = 0.2$ dB/km, $\beta_2 = -20.87$ ps$^2$/km, $\gamma = 1.3$ rad/W/km), where EDFAs (noise figure $5.0$ dB) compensate for attenuation after each span. Forward propagation is simulated with $300$ logarithmic StPS and $100$ GHz simulation bandwidth. No PMD or other hardware impairments are included in the simulations. At the receiver, the signal is low-pass filtered ($30$ GHz bandwidth) and downsampled to $2$ samples/symbol for further processing. LDBP is applied first, followed by a matched filter\footnote{For the experimental setup, matched filtering is implicitly performed by the MIMO equalizer.} (MF) and phase-offset correction. LDBP is based on the symmetric SSFM using $3$ StPS and a logarithmic step size. When combining the adjacent linear half-steps, the overall model has $61$ linear steps. MSE \RevA{is the loss function} employed for training all FIR filters, \RevA{defined as $\sum_{p\in \{\text{x},\text{y}\}} \lVert \vect{y}_p - \hat{\vect{y}}_p \rVert^2/2$, where  $\vect{y}_p$ and $\hat{\vect{y}}_p$ are the transmitted and estimated symbol vectors of the $p$-polarization after the phase-offset correction, respectively.} We assume that \RevB{the filters are symmetric and that} different filters are used in each polarization. This is essentially the same methodology as described in earlier work \cite{Haeger2018ofc, Haeger2018isit}.

Compared to most prior work on complexity-reduced DBP, it should be stressed that our goal is not to reduce the number of steps, but instead to reduce the per-step complexity. This is accomplished by employing filter pruning. All FIR filters are initialized with constrained least-squares CD coefficients according to \cite{Sheikh2016}. \RevA{The approach in \cite{Sheikh2016} minimizes the frequency-response error of the FIR filter with respect to an ideal CD compensation filter within the signal bandwidth, while constraining the out-of-band filter gain.} The initial filter lengths are chosen large enough to ensure good performance. The filters are then progressively pruned to a given target length by forcing the outermost taps to zero at certain iterations during SGD \cite{Fougstedt2018ecoc}. \RevA{The zero forcing is done using a masking operation in TensorFlow. The iterations where pruning occurs are predefined before the training begins.} For the considered scenario, the targeted model consisted of $22$ filters with $7$ taps and $39$ filters with $9$ taps. Training is performed for $50000$ iterations using the Adam optimizer \cite{Kingma2014}, learning rate $0.0007$, and batch size $50$\RevB{, which took around two hours on our machine. In principle, the number of iterations (and, hence, the training time) could be reduced, for example by setting a more aggressive learning rate. However, we observed that larger learning rates can sometimes lead to diverging MSE losses and numerical instabilities in our implementation.} \RevB{The filters are trained considering data from different launch powers, randomly chosen from the set $\mathcal{P} = \{1, 1.5, 2, 2.5, 3\}\,$dBm, resulting in a single model that tolerates changes in the input power. }



\subsection{Fine-Tuning with Experimental Data}
\label{sc:finet}

The next step is to fine-tune the pre-trained and pruned LDBP model using experimental data traces. The key challenge when training with experimental data is the presence of various hardware impairments and time-varying effects such as PMD and carrier phase noise. Our approach is to first estimate these impairments using the conventional DSP chain \NoRev{\Removed{(assuming EDC only)}} and then properly pre-process the data. The actual training is then performed with the \NoRev{resulting} pre-processed data. In particular: 

\begin{itemize}
    \item The received data samples are pre-processed by applying \NoRev{\Removed{the}} receiver front-end compensation and \NoRev{\Removed{the}} frequency-offset compensation. The frequency offset is estimated from the standard DSP chain. \NoRev{DBP is then applied to the resulting signal to improve the estimation of phase noise and the SOP. }
    
    \item The data symbols used for supervised learning are circular-shifted for alignment with the data samples.\footnote{\RevB{In principle, LDBP with asymmetric FIR filters could learn to recover a circular shift. However, due to the use of symmetric filters, a manual shift has to be applied. For the pre-training in Sec.~\ref{sc:prefp}, no circular shift is necessary since the sequences are already perfectly aligned.}} These data symbols are pre-processed using the estimated phase noise process to de-rotate the symbols. More precisely, let $e^{\imag \hat{\phi}_i}$ for $i = 1, \ldots, N$ be the estimated phase noise process, where $N$ is the length of the data trace. Then, training is performed using the pre-processed symbols $e^{- \imag \hat{\phi}_i} x_i$, $i = 1, \ldots, N$, where $x_i$ are the true data symbols.
    
    \item Finally, the filter taps for the adaptive MIMO equalizer after the first $60000$ symbols are extracted and saved for each data trace. The equalizer is then assumed to be static for the rest of the trace. During LDBP training, the MIMO equalizer is integrated as a static DSP component after the MF, where the saved filter coefficients are loaded and applied. Note that the MIMO filter taps are not updated during the fine tuning of LDBP. 
    
\end{itemize}

Fine-tuning using the above approach is performed for an additional $5000$ iterations using a learning rate $0.0007$, and batch size $50$. \NoRev{For training, we only consider 19 out of the 37 available traces for each launch power in the set $\mathcal{P}$, where the remaining traces are reserved for testing. }

\subsection{Testing}

\begin{figure}
    \centering
    \begin{tikzpicture}
\pgfplotsset{
legend image code/.code={
\draw[mark repeat=2,mark phase=2]
plot coordinates {
(0cm,0cm)
(0.2cm,0cm)        
(0.4cm,0cm)         
};}}
\begin{axis}[
xlabel = Transmitted power $P$ (dBm),
ylabel = Effective SNR (dB),
xmin=-5,
xmax=6,
ymin=15.5,
ymax=21.5,
legend style={at={(0.01,0.99)},anchor=north west, nodes={scale=0.85, transform shape}, font=\footnotesize},
legend cell align={left},
grid = major,
ytick distance = 1,
xtick distance = 1,
every axis plot/.append style={thick, each nth point=1},
ylabel near ticks,
xlabel near ticks,
]
\pgfplotstableread{tikz/avg_SNR_PCG_independent.txt}\mydata
\addplot[color = red, circle marker] table[x = Power, y = EDC]{\mydata};\addlegendentry{EDC};
\addplot[color = blue, square marker] table[x = Power, y = DBP_1_optimized]{\mydata};\addlegendentry{DBP $1$ StPS (freq. domain)};
\addplot[color = blue, diamond marker] table[x = Power, y = DBP_3_optimized]{\mydata};\addlegendentry{DBP $3$ StPS (freq. domain)};
\addplot[color = green!50!black, square marker, dashed] table[x = Power, y = LDBP_1_445_train]{\mydata};\addlegendentry{LDBP $1$ StPS ($445$ taps)};
\addplot[color = green!50!black, diamond marker, dashed] table[x = Power, y = LDBP_3_symmetric_even]{\mydata};\addlegendentry{LDBP $3$ StPS ($445$ taps)};
\addplot[color = green!50!black, dotted] table[x = Power, y = LDBP_3_symmetric_pruned]{\mydata};\addlegendentry{LDBP $3$ StPS ($323$ taps)}; 

\node[ellipse,semithick,minimum height=.65cm,minimum width=.65cm,draw] at (2.25,19.145) (ell1) {};
\draw[semithick] (ell1) -- (3.5,20.0);

\draw[thick,dashed] (-1.5,1.715282e+01) -- (2.65,1.715282e+01);
\draw[thick,<->] (1,1.715282e+01) -- (1,1.867241e+01)node[pos=0,above left = 12pt and -2pt]{\scriptsize 1.5 dB};
\draw[thick,<->] (2.5,1.715282e+01) -- (2.5,1.927653e+01)node[pos=0,above left = 12pt and -2pt]{\scriptsize 2.1 dB};

\draw[thick,dashed] (1,1.867241e+01) -- (-2,1.867241e+01);
\draw[thick,dashed] (2,1.901958e+01) -- (-2,1.901958e+01);
\draw[<->] (-1.75,1.867241e+01) -- (-1.75,1.901958e+01)node[pos=0,above left = -1.35pt and -1pt]{\scriptsize 0.35 dB};

\end{axis}

\begin{axis}[
every tick label/.append style={font=\footnotesize},
every axis plot/.append style={thick},
at={(5.05cm,4.2cm)},
width=3.3cm,
height=3cm,
ytick={18.,19.0,19.2},
xtick=\empty,
xmin = 1.9, 
xmax = 2.6,
ymin = 18.9, 
ymax = 19.35,
axis background/.style={fill=white}]
\pgfplotstableread{tikz/avg_SNR_PCG_independent.txt}\mydata
\addplot[color = blue, diamond marker] table[x = Power, y = DBP_3_optimized]{\mydata};
\addplot[color = green!50!black, square marker, dashed] table[x = Power, y = LDBP_1_445_train]{\mydata};
\addplot[color = green!50!black, diamond marker, dashed] table[x = Power, y = LDBP_3_symmetric_even]{\mydata};
\addplot[color = green!50!black, dotted] table[x = Power, y = LDBP_3_symmetric_pruned]{\mydata};

\draw[thick,dashed] (2,1.896964e+01) -- (2.3,1.896964e+01);

\draw[thick,dashed] (2.5,1.929453e+01)--(2.1,1.929274e+01);

\draw[<->,thick] (2.2,1.929453e+01) -- (2.2,1.896964e+01)node[right,pos=0.5]{\scriptsize 0.32 dB};

\end{axis}

\end{tikzpicture}
    \vspace{-0.2cm}
    \caption{Experimental results for single-channel transmission of a \SI{25}{Gbaud} signal with PM $16$-QAM over \SI{1500}{km}. }
    \vspace{-0.2cm}
    \label{fig:results}
\end{figure}
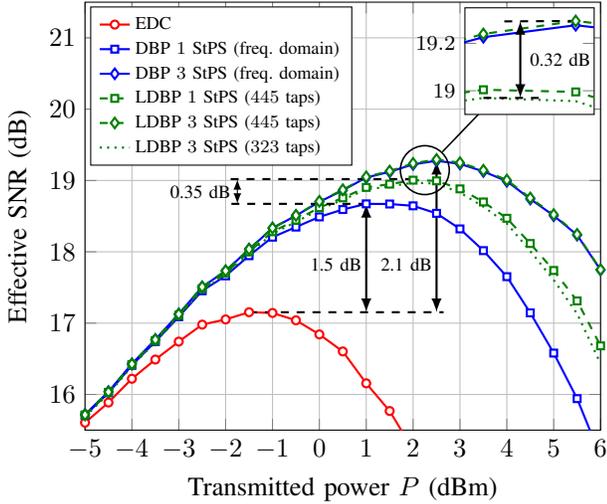

After fine-tuning is completed, the obtained model can then be used as a static nonlinear equalizer in the standard DSP chain described in Sec.~\ref{sc:dsp} (see middle block in Fig.~\ref{fig:dspchain}). During the testing phase, all DSP blocks are operated normally and the raw (i.e., not pre-processed) experimental data is used. The obtained performance is shown in Fig.~\ref{fig:results} \NoRev{with circle and diamond markers} for EDC, standard DBP \NoRev{($3$ StPS)}, and LDBP \NoRev{($3$ StPS)}. For DBP, a well-known issue is that the nonlinearity parameter $\gamma$ is usually not known precisely and needs to be estimated \cite{Lin2014}. We performed a simple grid search over $\gamma$\RevB{, jointly with $\beta_2$,} in order to optimize performance \RevB{\Removed{at $1$ dBm launch power}. For DBP with 3 StPS, the optimization was done at $2.5$ dBm launch power, }which led to \RevA{an} optimal value of \RevB{\Removed{$\gamma = 1.06$ rad/W/km}} \RevB{$\gamma = 1.21$ rad/W/km and $\beta_2 = -20.90$ ps$^2$/km. The latter is close to the experimentally estimated $\beta_2 = -20.87$ ps$^2$/km.} \RevB{Similar to $\gamma$ and $\beta_2$, the fiber length is also usually not precisely determined. We already had an estimated measure of $75.484$ km for the span length, which was confirmed to be optimum after a grid search. The optimum attenuation coefficient for DBP was $\alpha=0.19$ dB/km, the same as in the fiber specifications.} DBP with $3$ StPS achieves a peak-SNR gain of \RevB{\Removed{$1.7$} $2.1$} dB over EDC. The peak-SNR gain obtained by DBP is similar to those reported in prior experimental studies on single-channel DBP \cite{Lin2015, Galdino2017}. DBP also improves the optimum launch power with respect to EDC by \RevB{\Removed{$3$} $4$} dB, from \RevB{\Removed{$-1$} $-1.5$} dBm to \RevB{\Removed{$2$} $2.5$} dBm. By using LDBP, the peak-SNR gain is \RevB{slighly} increased \RevB{\Removed{by $0.2$ dB}} with respect to DBP. \RevB{\Removed{and $1.9$ dB with respect to EDC}} Further increasing the number of StPS \RevA{or filter taps for LDBP} did not improve performance. The optimum launch power for LDBP remains the same as the one for DBP. \NoRev{We also repeated the same procedure assuming $1$ StPS for both DBP and LDBP, in which case the LDBP model uses $21$ filters per polarization, where $12$ filters are pruned to $23$ taps and $9$ filters to $21$ taps. The results in Fig.~\ref{fig:results} (square markers) show that in this case LDBP achieves a performance improvement of around $0.35$ dB with respect to DBP.} \RevB{For DBP with 1 StPS, the optimum values for $\gamma$ and $\beta_2$ were found to be $\gamma = 1.62$ rad/W/km and $\beta_2 = -21.41$ ps$^2$/km.} \RevB{\Removed{These results indicate that LDBP can learn to compensate for experimental impairments and other imperfections that are not directly accounted for by standard DBP.}}






\begin{figure*}[t]
	\centering
	\includegraphics{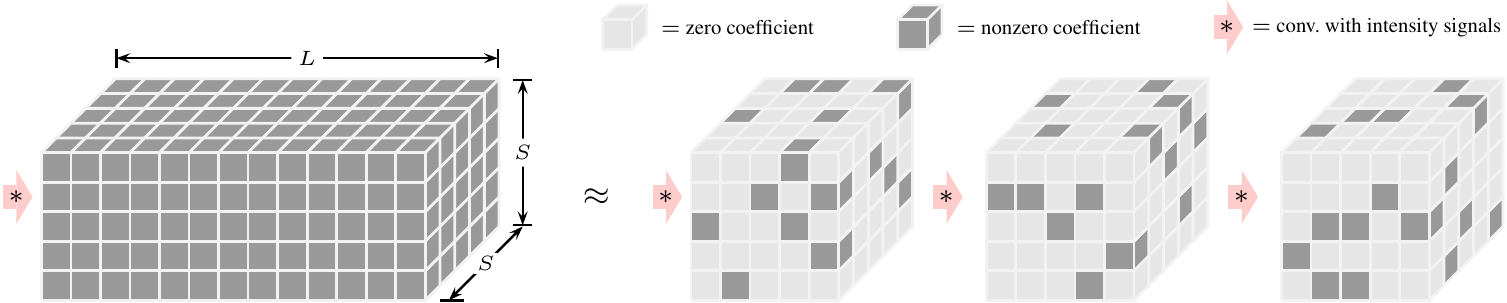}
	\caption{Tensor representation of an $L$-tap $S \times S$
	MIMO filter for DBP based on subband processing, where $S$ is the
	number of subbands (left); learned multi-step decomposition with
	sparse subfilters (right). }
	\label{fig:subbands}
\end{figure*}

In terms of complexity, \RevA{it has been shown that the power consumption and chip area for time-domain DBP \cite{Fougstedt2018ptl} and LDBP \cite{Fougstedt2018ecoc} are dominated by the linear steps, whereas the nonlinear steps have efficient hardware implementations using a Taylor expansion. Therefore, we focus on the linear steps for simplicity. As a simple surrogate measure for complexity, we use the overall impulse response length of the entire LDBP model, which is defined as the length of the filter obtained by convolving all LDBP subfilters. Since the same filter lengths are used in both polarizations, one may focus on a single polarization. For the $3$-StPS model, we have $22$ filters of length $7$ and $39$ filters of length $9$. Hence, the overall impulse response length is $2(22 \cdot (7-1)/2 + 39 \cdot (9-1)/2)+1=445$ taps. For the $1$-StPS model, the overall impulse response length is $2(12 \cdot (23-1)/2 + 9 \cdot (21-1)/2)+1=445$, i.e., the same as the $3$-StPS model.} \NoRev{Thus, even though the number of steps is reduced by a factor of $3$ and performance decreases by around $0.3$ dB (see the inlet figure in Fig.~\ref{fig:results}), the expected hardware complexity of the two models is roughly comparable. Moreover, the overall impulse response lengths} should be compared to the memory that is introduced by CD. To estimate the memory, one may use the fact that CD leads to a group delay difference of $2 \pi |\beta_2| \Delta f L_{\text{tot}}$ over a bandwidth $\Delta f$ and transmission distance $L_{\text{tot}}$. Normalizing by the sampling interval $T$, this confines the memory to roughly $(2 \pi |\beta_2| \Delta f L_{\text{tot}})/T$ samples. For our scenario, we have $\beta_2 = -20.87$ ps$^2$/km, $L_{\text{tot}} \approx 1510$ km, and $1/T = 50$ GHz. The bandwidth $\Delta f$ depends on the baud rate, the pulse shaping filter, and the spectral broadening during propagation. \RevA{\Removed{We set $\Delta f$ to be $60$\% of the DSP bandwidth $1/T$.} In order to obtain an estimate for $\Delta f$, we quantified the effect of spectral broadening in an ideal noiseless simulation environment for the same parameters as listed in Sec.~\ref{sc:prefp}. The bandwidth percentage (with respect to $1/T$) that contained $99.9$\% of the received signal power was found to vary between $51$\% for $P = -4\,$dBm and $77$\% for $P=6\,$dBm.} With these numbers, the CD memory \RevA{varies between $253$ and $381$} taps, which is comparable to the impulse response length of LDBP. This is a major improvement compared to previous work where the filter lengths in DBP are significantly longer than the CD memory, sometimes by orders of magnitude \cite{Ip2008, Martins2018}. We also note that it is possible to further prune the filters at the expense of some performance loss. \RevA{To illustrate this, we further pruned the 3-StPS model to $22$ filters of length $5$ and $39$ filters of length $7$. This gave a peak-SNR penalty of $0.32$ dB (see Fig.~\ref{fig:results}), making the performance comparable to the $1$-StPS model, while at the same time reducing the overall impulse response length to only $323$ taps.}



\section{Outlook and Future Work} 
\label{sc:outlook}

In this section, we give an overview of related work on LDBP that has previously appeared in the literature and also comment on potentially interesting avenues for future work. 

\subsection{Sparse MIMO Filters for Subband Processing}

The complexity of DBP with TD filtering is largely dominated by the total number of required CD filter taps in all steps and this increases quadratically with bandwidth, see, e.g., \cite{Taylor2008, Ho2009}. Thus, efficient TD-DBP of wideband signals is challenging. One possible solution is to employ subband processing and split the received signal into $S$ parallel signals using a filter bank \cite{Taylor2008, Ho2009, Slim2013, Nazarathy2014, Mateo2010, Ip2011, Oyama2015, Haeger2018ecoc}. A theoretical foundation for DBP based on subband processing is obtained by inserting the split-signal assumption $u = \sum_{i=1}^S u_i$ into the NLSE. This leads to a set
of coupled equations which can then be solved numerically. We focus on
the modified SSFM proposed in \cite{Leibrich2003} which is essentially
equivalent to the standard SSFM for each subband, except that all
sampled intensity waveforms $|u_1|^2, \ldots, |u_S|^2$ are jointly
processed with a MIMO filter prior to each nonlinear phase rotation
step. This accounts for cross-phase modulation between subbands but
not four-wave mixing because no phase information is exchanged. 

The MIMO filters for subband processing can be relatively demanding in terms of hardware complexity. As an example, in
\cite{Haeger2018ecoc} we considered a scenario where a $96$-Gbaud signal is split into $S=7$
subbands. For a filter length of $13$, the MIMO filter in each SSFM
step can be represented as a $7 \times 7 \times 13$ tensor with $637$
real coefficients which is shown in Fig.~\ref{fig:subbands} (left).
The resulting complexity per step and subband would be almost $6$
times larger than that of the CD filters used in
\cite{Haeger2018ecoc}. The situation can be improved significantly by
decomposing each MIMO filter into a cascade of sparse filters as shown
in Fig.~\ref{fig:subbands} (right). For a cascade of $3$ filters, it
was shown that a simple $L_1$-norm regularization applied to the
filter coefficients during SGD leads to a sparsity level of round
$8$\%, i.e., $92$\% of the filter coefficients can be set to zero with
little performance penalty. Note that this filter decomposition
happens \emph{within} each SSFM step. In other words, complexity is
reduced by further increasing the depth of the multi-step DBP
computation graph. 

\subsection{Distributed PMD Compensation}
\label{sec:pmd}

Different techniques have been proposed in previous works to embed the distributed compensation of PMD in the DBP algorithm, when the knowledge of the PMD evolution in the link is missing \cite{Liga2018, Czegledi2017, Goroshko2016}. In this section, we describe how distributed PMD compensation can be combined with LDBP in a hardware-efficient manner. 

\begin{figure*}[t]
	\centering
	\includegraphics{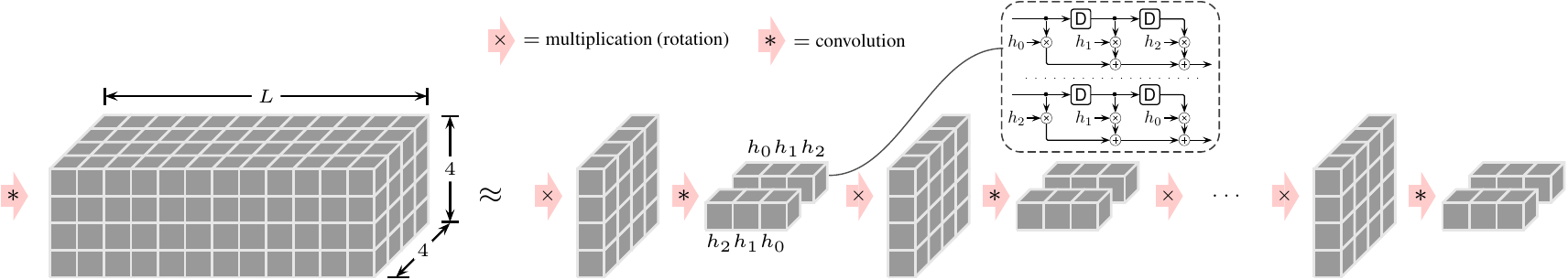}
	\caption{Tensor representation of an $L$-tap $4 \times 4$ MIMO filter for
	PMD compensation (left); multi-step decomposition where
	$4$-D rotations are alternated with short fractional-delay (FD)
	filters accounting for DGD (right). Each FD filters is applied to
	both the real and imaginary part. }
	\label{fig:pmd_tensor}
\end{figure*}

As discussed in Sec.~\ref{sc:fiberprop}, PMD can be modeled by dividing a fiber link of length $L_{\text{tot}}$ into $M=L_{\text{tot}}/h$ sections, where for large enough $M$ the link Jones matrix $\bJ_{\text{Link}}(\omega)$ can be factorized as  
\begin{equation}
\label{eq:dgd}
\mat{J}_{\text{Link}}(\omega)\triangleq
\exp\left(-j\omega\int_{0}^{L_{\text{tot}}}\Delta \beta '(\xi)\overline{\psig}(\xi)d\xi\right)=\prod_{i=1}^M \mat{R}^{(i)} \mat{T}^{(i)}(\omega)
\end{equation}
where $\mat{R}^{(i)}\triangleq \mat{R}(ih)$ and  $\mat{T}^{(i)}(\omega)\triangleq \mat{T}(ih,\omega)$ for $i=1,2,\ldots,M$. PMD compensation (and polarization demultiplexing) then
amounts to finding and applying the inverse $\mat{J}_{\text{Link}}^{-1}(\omega)$ to
the received signal. 
This is typically performed after CD compensation, e.g., using an $L$-tap MIMO filter that tries to approximate $\mat{J}_{\text{Link}}^{-1}(\omega)$. Fig.~\ref{fig:pmd_tensor} (left) shows the corresponding tensor representation assuming a real-valued $4 \times 4$ filter that is applied to the separated real and imaginary parts of both polarizations \cite{Crivelli2014}. 

An efficient multi-step decomposition of this filter is depicted in Fig.~\ref{fig:pmd_tensor} (right), which essentially mimics \eqref{eq:dgd} in a reverse fashion. Here, the matrices $\mat{T}^{(i)}(\omega)$ are approximated with two real-valued fractional-delay (FD) filters employing symmetrically flipped filter coefficients for different polarizations. The FD filters can be very short provided that the expected DGD per step is sufficiently small (i.e., many steps are used). In \cite{Haeger2020ofc}, it was shown how to integrate the decomposed filter structure into LDBP. The resulting multi-step PMD architecture can be trained effectively using SGD. An important feature compared to previous work is the fact that the employed approach does not assume any knowledge about the particular PMD realizations along the link, nor any knowledge about the total accumulated PMD. However, more research is needed to fully characterize the training behavior, e.g., in terms of convergence speed for adaptive compensation.

\subsection{Coefficient Quantization and ASIC Implementation}

Fixed-point requirements and other DSP hardware implementation aspects
for DBP have been investigated in \cite{Fougstedt2017,
Fougstedt2018ecoc, Fougstedt2018ptl, Martins2018, Sherborne2018}. A
potential benefit of multi-step architectures is that they empirically
tend to have many ``good'' parameter configurations that lie
relatively close to each other. This implies that even if the
optimized parameters are slightly perturbed (e.g., by quantizing them)
there may exist a nearby parameter configuration that exhibits
similarly good performance to mitigate the resulting performance loss
due to the perturbation. 

Numerical evidence for this phenomenon can be obtained by considering
the joint optimization of CD filters in DBP including the effect of
filter coefficient quantization. This has been studied in
\cite{Fougstedt2018ecoc} and the approach relies on applying so-called
``fake'' quantizations to the filter coefficients, where the gradient
computations and parameter updates during SGD are still performed in
floating point. Compared to other quantization methods, this jointly
optimizes the responses of quantized filters and can lead to
significantly reduced fixed-point requirements. For the scenario in
\cite{Fougstedt2018ecoc}, it was shown for example that the bit
resolution can be reduced from $8$-$9$ coefficient bits to $5$--$6$
bits without adversely affecting performance. Furthermore, hardware
synthesis results in $28$-nm CMOS show that multi-step DBP based on TD
filtering with short FIR filters is well within the limits of current
ASIC technology in terms of chip area and power consumption
\cite{Fougstedt2018ecoc, Fougstedt2018ptl}.

\section{Conclusions}\label{sc:conc}

We have illustrated how machine learning can be used to achieve efficient fiber-nonlinearity compensation. Rather than reducing the number of steps (or steps per span), it was highlighted that complexity can also be reduced by carefully designing and optimizing multi-step methods, or even by increasing the number of steps and decomposing complex operations into simpler ones, without losing performance. We also avoided the use of neural networks as universal (but sometimes poorly understood) function approximators. Instead, the considered learned digital backpropagation relied on parameterizing the split-step method, i.e., an existing model-based algorithm. We have performed an experimental demonstration of this approach, which was shown to outperform standard digital backropagation with limited complexity. Some extensions of the approach and steps towards possible future works were also presented, showing that there is a possibility for further performance improvements in these systems. 

\section*{Acknowledgments}

This work is part of a project that has received funding from the
European Union's Horizon 2020 research and innovation programme under
the Marie Sk\l{}odowska-Curie grant agreement No.~749798. The work of
H.~D.~Pfister was supported in part by the National Science Foundation
(NSF) under Grant No.~1609327. The work of A.~Alvarado, G.~Liga, and S. Goossens
has received funding from the European Research Council (ERC) under
the European Union's Horizon 2020 research and innovation programme
(grant agreement No.~757791). The work of G. Liga is also funded by the EUROTECH postdoc programme under the European Union’s Horizon 2020 research and innovation programme (Marie Skłodowska-Curie grant agreement No 754462). C. Okonkwo and S. van der Heide are partially funded by Netherlands Organisation for Scientific Research (NWO) Gravitation Program on Research Center for Integrated Nanophotonics (GA 024.002.033). This work is also supported by the NWO via the VIDI Grant ICONIC (project number 15685). Any opinions, findings, recommendations, and conclusions expressed in this material are those
of the authors and do not necessarily reflect the views of these
sponsors.

\balance
\ifCLASSOPTIONcaptionsoff
  \newpage
\fi

\bibliographystyle{clsbibfiles/IEEEtran}
\bibliography{clsbibfiles/IEEEabrv,clsbibfiles/Bibliography}

\begin{thebibliography}{10}
\providecommand{\url}[1]{#1}
\csname url@rmstyle\endcsname
\providecommand{\newblock}{\relax}
\providecommand{\bibinfo}[2]{#2}
\providecommand\BIBentrySTDinterwordspacing{\spaceskip=0pt\relax}
\providecommand\BIBentryALTinterwordstretchfactor{4}
\providecommand\BIBentryALTinterwordspacing{\spaceskip=\fontdimen2\font plus
\BIBentryALTinterwordstretchfactor\fontdimen3\font minus
  \fontdimen4\font\relax}
\providecommand\BIBforeignlanguage[2]{{%
\expandafter\ifx\csname l@#1\endcsname\relax
\typeout{** WARNING: IEEEtran.bst: No hyphenation pattern has been}%
\typeout{** loaded for the language `#1'. Using the pattern for}%
\typeout{** the default language instead.}%
\else
\language=\csname l@#1\endcsname
\fi
#2}}

\bibitem{Agrell2016}
E.~Agrell, A.~Alvarado, and F.~R. Kschischang, ``{Implications of information
  theory in optical fibre communications},'' \emph{Philosophical Transactions
  of the Royal Society A: Mathematical, Physical and Engineering Sciences},
  vol. 374, Mar. 2016.

\bibitem{AGRAWAL201327}
G.~Agrawal, \emph{Nonlinear Fiber Optics}, 5th~ed., ser. Optics and
  Photonics.\hskip 1em plus 0.5em minus 0.4em\relax Boston: Academic Press,
  2013.

\bibitem{Li2008}
X.~Li, \emph{et~al.}, ``{Electronic post-compensation of {WDM} transmission
  impairments using coherent detection and digital signal processing},''
  \emph{Opt. Express}, vol.~16, no.~2, pp. 880--888, Jan. 2008.

\bibitem{Mateo2008}
E.~Mateo, L.~Zhu, and G.~Li, ``{Impact of {XPM} and {FWM} on the digital
  implementation of impairment compensation for {WDM} transmission using
  backward propagation},'' \emph{Opt. Express}, vol.~16, no.~20, pp.
  16\,124--16\,137, Sept. 2008.

\bibitem{Ip2008}
E.~Ip and J.~M. Kahn, ``{Compensation of dispersion and nonlinear impairments
  using digital backpropagation},'' \emph{J. Lightw. Technol.}, vol.~26,
  no.~20, pp. 3416--3425, Oct. 2008.

\bibitem{Millar2010}
D.~S. Millar, \emph{et~al.}, ``{Mitigation of fiber nonlinearity using a
  digital coherent receiver},'' \emph{IEEE J. Sel. Topics. Quantum Electron.},
  vol.~16, no.~5, pp. 1217--1226, Sept. 2010.

\bibitem{Peddanarappagari1997}
K.~V. Peddanarappagari and M.~Brandt-Pearce, ``{Volterra series transfer
  function of single-mode fibers},'' \emph{J. Lightw. Technol.}, vol.~15,
  no.~12, pp. 2232--2241, Dec. 1997.

\bibitem{Gao2009a}
Y.~Gao, F.~Zhang, L.~Dou, Z.~Chen, and A.~Xu, ``{Intra-channel nonlinearities
  mitigation in pseudo-linear coherent QPSK transmission systems via nonlinear
  electrical equalizer},'' \emph{Opt. Communications}, vol. 282, no.~12, pp.
  2421--2425, 2009.

\bibitem{Liu2012}
L.~Liu, \emph{et~al.}, ``Intrachannel nonlinearity compensation by inverse
  {Volterra} series transfer function,'' \emph{J. Lightw. Technol.}, vol.~30,
  no.~3, pp. 310--316, Feb. 2012.

\bibitem{Guiomar2012}
F.~P. Guiomar, J.~D. Reis, A.~L. Teixeira, and A.~N. Pinto, ``{Mitigation of
  intra-channel nonlinearities using a frequency-domain {Volterra} series
  equalizer},'' \emph{Opt. Express}, vol.~20, no.~2, pp. 1360--1369, Jan. 2012.

\bibitem{Yan2011}
W.~{Yan}, \emph{et~al.}, ``Low complexity digital perturbation
  back-propagation,'' in \emph{Proc. European Conf. Optical Communication
  (ECOC)}, Geneva, Switzerland, Sept. 2011.

\bibitem{Tao2011}
Z.~Tao, L.~Dou, W.~Yan, L.~Li, T.~Hoshida, and J.~C. Rasmussen,
  ``{Multiplier-free intrachannel nonlinearity compensating algorithm operating
  at symbol rate},'' \emph{J. Lightw. Technol.}, vol.~29, no.~17, pp.
  2570--2576, Sept. 2011.

\bibitem{Liang2014}
X.~Liang and S.~Kumar, ``{Multi-stage perturbation theory for compensating
  intra-channel nonlinear impairments in fiber-optic links},'' \emph{Opt.
  Express}, vol.~22, no.~24, p. 29733, Dec. 2014.

\bibitem{Nakashima2017}
H.~Nakashima, T.~Oyama, C.~Ohshima, Y.~Akiyama, Z.~Tao, and T.~Hoshida,
  ``{Digital nonlinear compensation technologies in coherent optical
  communication systems},'' in \emph{Proc. Optical Fiber Communication Conf.
  (OFC)}, Los Angeles, CA, 2017.

\bibitem{Cartledge:17}
J.~C. Cartledge, F.~P. Guiomar, F.~R. Kschischang, G.~Liga, and M.~P. Yankov,
  ``Digital signal processing for fiber nonlinearities,'' \emph{Opt. Express},
  vol.~25, no.~3, pp. 1916--1936, Feb. 2017.

\bibitem{Du2010}
L.~B. Du and A.~J. Lowery, ``{Improved single channel backpropagation for
  intra-channel fiber nonlinearity compensation in long-haul optical
  communication systems.}'' \emph{Opt. Express}, vol.~18, no.~16, pp.
  17\,075--17\,088, July 2010.

\bibitem{Rafique2011a}
D.~Rafique, M.~Mussolin, M.~Forzati, J.~M{\aa}rtensson, M.~N. Chugtai, and
  A.~D. Ellis, ``{Compensation of intra-channel nonlinear fibre impairments
  using simplified digital back-propagation algorithm.}'' \emph{Opt. Express},
  vol.~19, no.~10, pp. 9453--9460, Apr. 2011.

\bibitem{Napoli2014}
A.~Napoli, \emph{et~al.}, ``{Reduced complexity digital back-propagation
  methods for optical communication systems},'' \emph{J. Lightw. Technol.},
  vol.~32, no.~7, pp. 1351--1362, Apr. 2014.

\bibitem{Jarajreh2015}
A.~M. Jarajreh, \emph{et~al.}, ``{Artificial neural network nonlinear equalizer
  for coherent optical {OFDM}},'' \emph{IEEE Photon. Technol. Lett.}, vol.~27,
  no.~4, pp. 387--390, Feb. 2015.

\bibitem{Giacoumidis2015}
E.~Giacoumidis, \emph{et~al.}, ``{Fiber nonlinearity-induced penalty reduction
  in {CO-OFDM} by {ANN}-based nonlinear equalization},'' \emph{Opt. Lett.},
  vol.~40, no.~21, pp. 5113--5116, Nov. 2015.

\bibitem{Secondini2016}
M.~Secondini, S.~Rommel, G.~Meloni, F.~Fresi, E.~Forestieri, and L.~Poti,
  ``{Single-step digital backpropagation for nonlinearity mitigation},''
  \emph{Photon. Netw. Commun.}, vol.~31, no.~3, pp. 493--502, 2016.

\bibitem{Fougstedt2017}
C.~Fougstedt, M.~Mazur, L.~Svensson, H.~Eliasson, M.~Karlsson, and
  P.~Larsson-Edefors, ``{Time-domain digital back propagation: algorithm and
  finite-precision implementation aspects},'' in \emph{Proc. Optical Fiber
  Communication Conf. (OFC)}, Los Angeles, CA, 2017.

\bibitem{Haeger2018ofc}
C.~H{\"{a}}ger and H.~D. Pfister, ``{Nonlinear interference mitigation via deep
  neural networks},'' in \emph{Proc. Optical Fiber Communication Conf. (OFC)},
  San Diego, CA, 2018.

\bibitem{Lin2017}
H.~W. Lin, M.~Tegmark, and D.~Rolnick, ``{Why does deep and cheap learning work
  so well?}'' \emph{J. Stat. Phys.}, vol. 168, no.~6, pp. 1223--1247, Sept.
  2017.

\bibitem{Haeger2018isit}
C.~H{\"{a}}ger and H.~D. Pfister, ``{Deep learning of the nonlinear
  {S}chr{\"{o}}dinger equation in fiber-optic communications},'' in \emph{Proc.
  IEEE Int. Symp. Information Theory (ISIT)}, Vail, CO, 2018.

\bibitem{Fougstedt2018ecoc}
C.~Fougstedt, C.~H{\"{a}}ger, L.~Svensson, H.~D. Pfister, and
  P.~Larsson-Edefors, ``{{ASIC} implementation of time-domain digital
  backpropagation with deep-learned chromatic dispersion filters},'' in
  \emph{Proc. European Conf. Optical Communication (ECOC)}, Rome, Italy, 2018.

\bibitem{Lecun1989}
Y.~Lecun, J.~S. Denker, and S.~A. Solla, ``{Optimal Brain Damage},'' in
  \emph{Proc. Advances in Neural Information Processing Systems (NIPS)},
  Denver, CO, 1989.

\bibitem{Han2016}
S.~Han, H.~Mao, and W.~J. Dally, ``{Deep compression: compressing deep neural
  networks with pruning, trained quantization and Huffman coding},'' in
  \emph{Proc. Int. Conf. Learning Representations (ICLR)}, San Juan, Puerto
  Rico, 2016.

\bibitem{Haeger2019ecoc}
C.~H{\"{a}}ger, H.~D. Pfister, R.~M. B{\"{u}}tler, G.~Liga, and A.~Alvarado,
  ``{Revisiting multi-step nonlinearity compensation with machine learning},''
  in \emph{Proc. European Conf. Optical Communication (ECOC)}, Dublin, Ireland,
  2019.

\bibitem{sillekens2019experimental}
E.~Sillekens, \emph{et~al.}, ``Experimental demonstration of learned
  time-domain digital back-propagation,'' \emph{arXiv:1912.12197}, Dec. 2019.

\bibitem{Bitachon2020}
B.~I. Bitachon, A.~Ghazisaeidi, B.~Baeuerle, M.~Eppenberger, and J.~Leuthold,
  ``{Deep Learning Based Digital Back Propagation with Polarization State
  Rotation \& Phase Noise Invariance},'' in \emph{Proc. Optical Fiber
  Communication Conf. (OFC)}, San Diego, CA, 2020.

\bibitem{Marcuse1997}
D.~Marcuse, C.~R. Menyuk, and P.~K.~A. Wai, ``{Application of the Manakov-PMD
  equation to studies of signal propagation in optical fibers with randomly
  varying birefringence},'' \emph{J. Lightw. Technol.}, vol.~15, no.~9, pp.
  1735--1745, Sept. 1997.

\bibitem{Ip2010}
E.~Ip, ``{Nonlinear compensation using backpropagation for
  polarization-multiplexed transmission},'' \emph{J. Lightw. Technol.},
  vol.~28, no.~6, pp. 939--951, Mar. 2010.

\bibitem{Gilmore1974}
R.~{Gilmore}, ``{Baker-Campbell-Hausdorff formulas},'' \emph{J. Mathematical
  Physics}, vol.~15, no.~12, pp. 2090--2092, Dec. 1974.

\bibitem{Sinkin2002}
O.~V. {Sinkin}, R.~{Holzlohner}, J.~{Zweck}, and C.~R. {Menyuk}, ``Optimization
  of the split-step {Fourier} method in modeling optical-fiber communications
  systems,'' \emph{J. Lightw. Technol.}, vol.~21, no.~1, pp. 61--68, Jan. 2003.

\bibitem{Czegledi2017}
C.~B. Czegledi, \emph{et~al.}, ``{Digital backpropagation accounting for
  polarization-mode dispersion},'' \emph{Opt. Express}, vol.~25, no.~3, pp.
  1903--1915, Feb. 2017.

\bibitem{Liga2018}
G.~Liga, C.~Czegledi, and P.~Bayvel, ``{A {PMD}-adaptive {DBP} receiver based
  on {SNR} optimization},'' in \emph{Proc. Optical Fiber Communication Conf.
  (OFC)}, San Diego, CA, 2018.

\bibitem{Savory2008}
S.~J. Savory, ``{Digital filters for coherent optical receivers},'' \emph{Opt.
  Express}, vol.~16, no.~2, pp. 804--817, Jan. 2008.

\bibitem{Zhu2009}
L.~Zhu, X.~Li, E.~Mateo, and G.~Li, ``{Complementary {FIR} filter pair for
  distributed impairment compensation of {WDM} fiber transmission},''
  \emph{IEEE Photon. Technol. Lett.}, vol.~21, no.~5, pp. 292--294, Mar. 2009.

\bibitem{Goldfarb2009}
G.~Goldfarb and G.~Li, ``{Efficient backward-propagation using wavelet- based
  filtering for fiber backward-propagation},'' \emph{Opt. Express}, vol.~17,
  no.~11, pp. 814--816, May 2009.

\bibitem{Lian2018itw}
M.~Lian, C.~H{\"{a}}ger, and H.~D. Pfister, ``{What can machine learning teach
  us about communications?}'' in \emph{Proc. IEEE Information Theory Workshop
  (ITW)}, Guangzhou, China, 2018.

\bibitem{pcg2014}
M.~E. O'Neill, ``Pcg: A family of simple fast space-efficient statistically
  good algorithms for random number generation,'' Harvey Mudd College,
  Claremont, CA, Tech. Rep. HMC-CS-2014-0905, Sept. 2014.

\bibitem{Sheikh2016}
A.~Sheikh, C.~Fougstedt, A.~{Graell i Amat}, P.~Johannisson,
  P.~Larsson-Edefors, and M.~Karlsson, ``{Dispersion compensation {FIR} filter
  with improved robustness to coefficient quantization errors},'' \emph{J.
  Lightw. Technol.}, vol.~34, no.~22, pp. 5110--5117, Nov. 2016.

\bibitem{Kingma2014}
D.~P. Kingma and J.~Ba, ``{Adam: a method for stochastic optimization},'' in
  \emph{Proc. Int. Conf. Learning Representations (ICLR)}, San Diego, CA, 2015.

\bibitem{Lin2014}
C.-Y. Lin, \emph{et~al.}, ``{Adaptive digital back-propagation for optical
  communication systems},'' in \emph{Proc. Optical Fiber Communication Conf.
  (OFC)}, San Franscisco, CA, 2014.

\bibitem{Lin2015}
C.~Lin, S.~Chandrasekhar, and P.~J. Winzer, ``{Experimental study of the limits
  of digital nonlinearity compensation in {DWDM} systems},'' in \emph{Proc.
  Optical Fiber Communication Conf. (OFC)}, Los Angeles, CA, 2015.

\bibitem{Galdino2017}
L.~Galdino, \emph{et~al.}, ``{On the limits of digital back-propagation in the
  presence of transceiver noise},'' \emph{Opt. Express}, vol.~25, no.~4, pp.
  4564--4578, Feb. 2017.

\bibitem{Fougstedt2018ptl}
C.~Fougstedt, L.~Svensson, M.~Mazur, M.~Karlsson, and P.~Larsson-Edefors,
  ``{{ASIC} implementation of time-domain digital back propagation for coherent
  receivers},'' \emph{IEEE Photon. Technol. Lett.}, vol.~30, no.~13, pp.
  1179--1182, July 2018.

\bibitem{Martins2018}
C.~S. Martins, L.~Bertignono, A.~Nespola, A.~Carena, F.~P. Guiomar, and A.~N.
  Pinto, ``{Efficient time-domain {DBP} using random step-size and multi-band
  quantization},'' in \emph{Proc. Optical Fiber Communication Conf. (OFC)}, San
  Diego, CA, 2018.

\bibitem{Taylor2008}
M.~G. Taylor, ``{Compact digital dispersion compensation algorithms},'' in
  \emph{Proc. Optical Fiber Communication Conf. (OFC)}, San Diego, CA, 2008.

\bibitem{Ho2009}
K.-P. Ho, ``{Subband equaliser for chromatic dispersion of optical fibre},''
  \emph{Electronics Lett.}, vol.~45, no.~24, pp. 1224--1226, Nov. 2009.

\bibitem{Slim2013}
I.~Slim, A.~Mezghani, L.~G. Baltar, J.~Qi, F.~N. Hauske, and J.~A. Nossek,
  ``{Delayed single-tap frequency-domain chromatic-dispersion compensation},''
  \emph{IEEE Photon. Technol. Lett.}, vol.~25, no.~2, pp. 167--170, Jan. 2013.

\bibitem{Nazarathy2014}
M.~Nazarathy and A.~Tolmachev, ``{Subbanded {DSP} architectures based on
  underdecimated filter banks for coherent {OFDM} receivers: Overview and
  recent advances},'' \emph{IEEE Signal Processing Mag.}, vol.~31, no.~2, pp.
  70--81, Mar. 2014.

\bibitem{Mateo2010}
E.~F. Mateo, F.~Yaman, and G.~Li, ``{Efficient compensation of inter-channel
  nonlinear effects via digital backward propagation in {WDM} optical
  transmission},'' \emph{Opt. Express}, vol.~18, no.~14, pp. 15\,144--15\,154,
  July 2010.

\bibitem{Ip2011}
E.~Ip, N.~Bai, and T.~Wang, ``{Complexity versus performance tradeoff for fiber
  nonlinearity compensation using frequency-shaped, multi-subband
  backpropagation},'' in \emph{Proc. Optical Fiber Communication Conf. (OFC)},
  Los Angeles, CA, 2011.

\bibitem{Oyama2015}
T.~Oyama, \emph{et~al.}, ``{Complexity reduction of perturbation-based
  nonlinear compensator by sub-band processing},'' in \emph{Proc. Optical Fiber
  Communication Conf. (OFC)}, Los Angeles, CA, 2015.

\bibitem{Haeger2018ecoc}
C.~H{\"{a}}ger and H.~D. Pfister, ``{Wideband time-domain digital
  backpropagation via subband processing and deep learning},'' in \emph{Proc.
  European Conf. Optical Communication (ECOC)}, Rome, Italy, 2018.

\bibitem{Leibrich2003}
J.~Leibrich and W.~Rosenkranz, ``{Efficient numerical simulation of
  multichannel {WDM} transmission systems limited by {XPM}},'' \emph{IEEE
  Photon. Technol. Lett.}, vol.~15, no.~3, pp. 395--397, Mar. 2003.

\bibitem{Goroshko2016}
K.~Goroshko, H.~Louchet, and A.~Richter, ``{Overcoming performance limitations
  of digital back propagation due to polarization mode dispersion},'' in
  \emph{Proc. Int. Conf. Transparent Optical Networks (ICTON)}, Trento, Italy,
  2016.

\bibitem{Crivelli2014}
D.~E. Crivelli, \emph{et~al.}, ``{Architecture of a single-chip 50 {Gb/s}
  {DP}-{QPSK/BPSK} transceiver with electronic dispersion compensation for
  coherent optical channels},'' \emph{IEEE Trans. Circuits Syst. I: Reg.
  Papers}, vol.~61, no.~4, pp. 1012--1025, Apr. 2014.

\bibitem{Haeger2020ofc}
C.~H{\"{a}}ger, H.~D. Pfister, R.~M. B{\"{u}}tler, G.~Liga, and A.~Alvarado,
  ``{Model-based machine learning for joint digital backpropagation and PMD
  compensation},'' in \emph{Invited paper at the Optical Fiber Communication
  Conf. {(OFC)}}, San Diego, CA, 2020.

\bibitem{Sherborne2018}
T.~Sherborne, B.~Banks, D.~Semrau, R.~I. Killey, P.~Bayvel, and D.~Lavery,
  ``{On the impact of fixed point hardware for optical fiber nonlinearity
  compensation algorithms},'' \emph{J. Lightw. Technol.}, vol.~36, no.~20, pp.
  5016--5022, Oct. 2018.

\end{thebibliography}

\vfill

\end{document}